# A Super-Polynomial Lower Bound for the Parity Game Strategy Improvement Algorithm as We Know it

Oliver Friedmann
Institut für Informatik, LMU München

E-mail: `Oliver.Friedmann@googlemail.com`

## Abstract

*This paper presents a new lower bound for the discrete strategy improvement algorithm for solving parity games due to Vöge and Jurdziński. First, we informally show which structures are difficult to solve for the algorithm. Second, we outline a family of games of quadratic size on which the algorithm requires exponentially many strategy iterations, answering in the negative the long-standing question whether this algorithm runs in polynomial time. Additionally we note that the same family of games can be used to prove a similar result w.r.t. the strategy improvement variant by Schewe.*

## 1. Introduction

Parity games are simple two-player games of perfect information played on directed graphs whose nodes are labeled with natural numbers, called priorities. A play in a parity game is an infinite sequence of nodes whose winner is determined by all priorities occurring infinitely often. In fact, it depends on the parity of the highest priority that occurs infinitely often, giving parity games their name.

Parity games occur in several fields of theoretical computer science, e.g. as solution to the problem of complementation or determinisation of tree automata [4, 1] or as algorithmic backend to the model checking problem of the modal $\mu$-calculus [2, 14].

There are many algorithms that solve parity games, such as the recursive decomposing algorithm due to Zielonka [17] and its recent improvement by Jurdziński, Paterson and Zwick [8], the small progress measures algorithm due to Jurdziński [7] with its recent improvement by Schewe [11], the model-checking algorithm due to Stevens and Stirling [13] and finally the two strategy improvement algorithms by Vöge and Jurdziński [16] and Schewe [12].

All mentioned algorithms except for the two strategy improvement algorithms have been shown to have a super-polynomial worst-case runtime complexity at best or there is at least little doubt that their worst-case runtime complexity is super-polynomial or even exponential.

Solving parity games is one of the few problems that belongs to the complexity class NP $\cap$ coNP and that is not (yet) known to belong to P [2]. It has also been shown that solving parity games belongs to UP $\cap$ coUP [6]. The currently best known upper bound on the deterministic solution of parity games is $\mathcal{O}(|E| \cdot |V|^{\frac{1}{3}|\mathtt{ran}\Omega|})$ due to Schewe's big-step algorithm [11].

In this paper, we present a family of parity games comprising a linear number of nodes and a quadratic number of edges such that the strategy improvement algorithm by Vöge and Jurdziński requires an exponential number of iterations on them. Consequently, the algorithm requires at least super-polynomial time to solve parity games in the worst case. Due to page restrictions, we will only study the original strategy improvement algorithm here, but we remark that the same result can be shown for the strategy improvement variant due to Schewe using the same family of games.

Section 2 defines the basic notions of parity games and some notations that are employed throughout the paper. Section 3 recaps the strategy improvement algorithm by Vöge and Jurdziński. In Section 4, we present two graph structures that are tricky to be solved by strategy iteration algorithms. Section 5 outlines a family of games on which the algorithm requires an exponential number of iterations.

## 2. Parity Games

A *parity game* is a tuple $G = (V, V_0, V_1, E, \Omega)$ where $(V, E)$ forms a directed graph whose node set is partitioned into $V = V_0 \cup V_1$ with $V_0 \cap V_1 = \emptyset$, and $\Omega : V \to \mathbb{N}$ is the

*priority function* that assigns to each node a natural number called the *priority* of the node. We assume the underlying graph to be total, i.e. for every $v \in V$ there is a $w \in W$ s.t. $(v, w) \in E$. In the following we will restrict ourselves to finite parity games.

We also use infix notation $vEw$ instead of $(v, w) \in E$ and define the set of all *successors* of $v$ as $vE := \{w \mid vEw\}$. The size $|G|$ of a parity game $G = (V, V_0, V_1, E, \Omega)$ is defined to be the cardinality of $E$, i.e. $|G| := |E|$; since we assume parity games to be total w.r.t. $E$, this seems to be a reasonable way to measure the size.

The game is played between two players called 0 and 1: Starting in a node $v_0 \in V$, they construct an infinite path through the graph as follows. If the construction so far has yielded a finite sequence $v_0 \ldots v_n$ and $v_n \in V_i$ then player $i$ selects a $w \in v_n E$ and the play continues with the sequence $v_0 \ldots v_n w$.

Every play has a unique winner given by the *parity* of the greatest priority that occurs infinitely often in a play. The winner of the play $v_0 v_1 v_2 \ldots$ is player $i$ iff $\max\{p \mid \forall j \in \mathbb{N} \exists k \geq j : \Omega(v_k) = p\} \equiv_2 i$ (where $i \equiv_k j$ holds iff $|i - j| \mod k = 0$). That is, player 0 tries to make an even priority occur infinitely often without any greater odd priorities occurring infinitely often, player 1 attempts the converse.

A graphical depiction of a parity game here is based on its directed graph where nodes owned by player 0 are drawn as circles and nodes owned by player 1 are drawn as rectangles; additionally, all nodes are labelled with their respective priority, and - if applicable - with their name.

A *strategy* for player $i$ is a function $\sigma : V_i \to V$, s.t. for all $v \in V_i$ holds that $vE\sigma(v)$. A play $v_0 v_1 \ldots$ *conforms* to a strategy $\sigma$ for player $i$ if for all $j$ we have: if $v_j \in V_i$ then $v_{j+1} = \sigma(v_j)$.

Intuitively, conforming to a strategy means to always make those choices that are prescribed by the strategy. A strategy $\sigma$ for player $i$ is a *winning strategy* starting in some node $v \in V$ if player $i$ wins every play that conforms to this strategy and begins in $v$. We say that player $i$ *wins* the game $G$ starting in $v$ iff the player has a winning strategy for $G$ starting in $v$.

With $G$ we associate two sets $W_0, W_1 \subseteq V$ with the following definition. $W_i$ is the set of all nodes $v$ s.t. player $i$ wins the game $G$ starting in $v$.

It is not obvious that every node should belong to either of $W_0$ or $W_1$. However, this is indeed the case and known as *determinacy*: a player has a strategy for a game iff the opponent does not have a strategy for that game.

**Theorem 1.** *[9, 5, 1] Let $G = (V, V_0, V_1, E, \Omega)$ be a parity game. Then $W_0 \cap W_1 = \emptyset$ and $W_0 \cup W_1 = V$.*

A strategy $\sigma$ for player $i$ induces a *strategy subgame* $G|_\sigma := (V, V_0, V_1, E|_\sigma, \Omega)$ where $E|_\sigma := \{(u, v) \in E \mid u \in dom(\sigma) \Rightarrow \sigma(u) = v\}$. Such a subgame $G|_\sigma$ is basically the same game as $G$ with the restriction that whenever $\sigma$ provides a strategy decision for a node $u \in V_i$ all transitions from $u$ but $\sigma(u)$ are no longer accessible. The set of strategies for player $i$ is denoted by $\mathcal{S}_i(G)$.

Without loss of generality we assume $\Omega$ to be injective, i.e. there are no two different nodes having the same priority. We also assume that parity games do not contain any self-cycles or if so that they are replaced by equivalent two-node cycles before passing the game to the strategy improvement algorithm.

## 3. Strategy Improvement

First, we briefly recap the basic definitions of the strategy improvement algorithm. For a given parity game $G = (V, V_0, V_1, E, \Omega)$, the *reward* of node $v$ is defined as follows: $\text{rew}_G(v) := \Omega(v)$ if $\Omega(v) \equiv_2 0$ and $\text{rew}_G(v) := -\Omega(v)$ otherwise. The set of *profitable nodes* for player 0 is defined to be $V_\oplus := \{v \in V \mid \text{rew}(v) \geq 0\}$ and $V_\ominus := \{v \in V \mid \text{rew}(v) < 0\}$ likewise for player 1.

The *relevance ordering* $<$ on $V$ is induced by $\Omega$: $v < u :\iff \Omega(v) < \Omega(u)$; additionally one defines the *reward ordering* $\prec$ on $V$ by $v \prec u :\iff \text{rew}_G(v) < \text{rew}_G(u)$. Note that both orderings are total due to injectivity of the priority function.

A *loopless path* in $G$ is an injective map $\pi : \{0, \ldots, k-1\} \to V$ conforming with $E$, i.e. $\pi(i) E \pi(i+1)$ for every $i < k$. The length of a loopless path is denoted by $|\pi| := k$. The set of loopless paths $\pi$ in a game $G$ originating from the node $v$ (i.e. $\pi(0) = v$) is denoted by $\Pi_G(v)$. We sometimes write $\pi = v_0 \ldots v_{k-1}$ to denote the loopless path $\pi : i \mapsto v_i$.

A node $v$ in $G$ is called *dominating cycle node* iff there is a loopless path $\pi \in \Pi_G(v)$ s.t. $\pi(|\pi| - 1) E \pi(0)$ and $\max\{\Omega(\pi(i)) \mid i < |\pi|\} = \Omega(v)$. The set of dominating cycles nodes is denoted by $\mathcal{C}_G$.

A key point of the strategy improvement algorithm is to assign to each node in the game graph a *valuation*. Basically, a valuation describes a loopless path originating from its node to a dominating cycle node. Such a valuation consists of three parts: The dominating cycle node, the set of more relevant nodes (w.r.t. the cycle node) on the loopless path leading to the cycle and the length of the loopless path (which measures the amount of less relevant nodes).

To compare the second component of a valuation - the set of nodes on the way to the cycle - we introduce a total ordering $\prec$ on $2^V$: To determine which set of nodes is better w.r.t. $\prec$, one investigates the node with the highest priority that occurs in only one of the two sets. The set owning that node is greater than the other if and only if that node has an even priority. More formally:

$$M \prec N :\iff$$
$$\begin{cases} (M \triangle N \neq \emptyset \wedge max_<(M \triangle N) \in N \cap V_\oplus) \vee \\ (M \triangle N \neq \emptyset \wedge max_<(M \triangle N) \in M \cap V_\ominus) \end{cases}$$

where $M \triangle N$ denotes the symmetric difference of both sets.

A loopless path $\pi = v_0 \ldots v_k$ induces a *node valuation* for the node $v_0$ as follows:

$$\vartheta_\pi := (v_k, \{v_i \mid v_k \prec v_i\}, k)$$

A node valuation $\vartheta$ for a node $v$ is a triple $(c, M, l) \in V \times 2^V \times |V|$ such that there is a loopless path $\pi$ with $\pi(0) = v$ and $\vartheta_\pi = \vartheta$.

We extend the total ordering on sets of nodes to node valuations:

$$(u, M, e) \prec (v, N, f) :\iff$$
$$\begin{cases} (u \prec v) \vee (u = v \wedge M \prec N) \vee \\ (u = v \wedge M = N \wedge e < f \wedge u \in V_\ominus) \vee \\ (u = v \wedge M = N \wedge e > f \wedge u \in V_\oplus) \end{cases}$$

A *game valuation* is a map $\Xi : V \to V \times 2^V \times |V|$ assigning each $v \in V$ a node valuation. A partial ordering on game valuations is defined as follows:

$$\Xi \triangleleft \Xi' :\iff (\forall v \in V : \Xi(v) \preceq \Xi'(v)) \wedge (\Xi \neq \Xi')$$

Game valuations are used to measure the performance of a strategy of player 0: For a fixed strategy $\sigma$ of player 0 and a node $v$, the associated valuation basically states which is the worst cycle that can be reached from $v$ conforming with $\sigma$ as well as the worst loopless path leading to that cycle (also conforming with $\sigma$). Intuitively, the associated valuation reflects the best counter-strategy player 1 could play.

A strategy $\sigma$ of player 0 therefore can be evaluated as follows:

$$\Xi_\sigma : v \mapsto \min_\prec \{\vartheta_\pi \mid \pi \in \Pi_{G|_\sigma}(v) \wedge \pi(|\pi| - 1) \in \mathcal{C}_{G|_\sigma}\}$$

**Lemma 2.** *[16] A valuation of a strategy can be computed in polynomial time.*

A valuation $\Xi$ originating from a strategy $\sigma$ can be used to create a new strategy of player 0. The strategy improvement algorithm only allows to select new strategy decisions for player 0 occurring in the *improvement arena* $\mathcal{A}_{G,\sigma} := (V, V_0, V_1, E', \Omega)$ where

$$vE'u :\iff$$
$$vEu \wedge (v \in V_1 \vee (v \in V_0 \wedge \Xi_\sigma(\sigma(v)) \preceq \Xi_\sigma(u)))$$

Thus all edges performing worse than the current strategy are removed from the game. A strategy $\sigma$ is *improvable* iff there is a node $v \in V_0$, a node $u \in V$ with $vEu$ and $\sigma(v) \neq u$ s.t. $\Xi_\sigma(\sigma(v)) \prec \Xi_\sigma(u)$.

An *improvement policy* now selects a strategy for player 0 in a given improvement arena w.r.t. a valuation originating from a strategy. More formally: An improvement policy is a map $\mathcal{I}_G : \mathcal{S}_0(G) \to \mathcal{S}_0(G)$ fulfilling the following two conditions for every strategy $\sigma$:

1. For every node $v \in V_0$ it holds that $(v, \mathcal{I}_G(\sigma)(v))$ is an edge in $\mathcal{A}_{G,\sigma}$.

2. If $\sigma$ is improvable then $\Xi_\sigma \neq \Xi_{\mathcal{I}_G(\sigma)}$.

Jurdziński and Vöge proved in their work that every strategy that is improved by an improvement policy can only result in strategies with valuations being better (w.r.t. $\triangleleft$) than the valuation of the original strategy.

**Theorem 3.** *[16] Let $G$ be a parity game, $\sigma$ be an improvable strategy and $\mathcal{I}_G$ be an improvement policy. Let $\sigma' = \mathcal{I}_G(\sigma)$. Then $\Xi_\sigma \triangleleft \Xi_{\sigma'}$.*

If a strategy is not improvable, the strategy iteration comes to an end.

**Theorem 4.** *[16] Let $G$ be a parity game and $\sigma$ be a non-improvable strategy. Then the following holds:*

1. $W_0 = \{v \mid \Xi_\sigma(v) = (w, \_, \_) \wedge w \in V_\oplus\}$

2. $W_1 = \{v \mid \Xi_\sigma(v) = (w, \_, \_) \wedge w \in V_\ominus\}$

3. $\sigma$ *is a winning strategy for player 0 on $W_0$*

4. $\tau : v \in V_1 \mapsto \min_\prec U_\sigma(v)$ *is a winning strategy for player 1 on $W_1$ where $U_\sigma(v) = \{u \in vE \mid \forall w \in vE : \Xi_\sigma(u) \preceq \Xi_\sigma(w)\}$.*

The strategy iteration starts with an initial strategy $\iota_G$ and runs for a given improvement policy $\mathcal{I}_G$ as follows.

---

**Algorithm 1** Strategy Iteration

1: $\sigma \leftarrow \iota_G$
2: **while** $\sigma$ is improvable **do**
3: $\quad \sigma \leftarrow \mathcal{I}_G(\sigma)$
4: **end while**
5: **return** $W_0, W_1, \sigma, \tau$ as in Theorem 4

---

The initial strategy can be selected in several ways. We focus on a very easy method here, always selecting the node with the best reward.

The initial strategy in this paper hence will be selected as follows:

$$\iota_G : v \in V_0 \mapsto \max_\prec \{u \mid vEu\}$$

The improvement policy we are following in this paper is the *locally optimizing policy* $\mathcal{I}_G^{\text{loc}}$ due to Jurdziński and Vöge.

It simply selects the most profitable strategy decision with respect to the current valuation:

$$\mathcal{I}_G^{\text{loc}}(\sigma): v \in V_0 \mapsto \max_{\prec} U_\sigma(v)$$

where $U_\sigma(v) = \{u \in vE \mid \forall w \in vE : \Xi_\sigma(w) \preceq \Xi_\sigma(u)\}$.

**Lemma 5.** *[16] The locally optimizing policy can be computed in polynomial time.*

We will also refer to the *globally optimizing policy* by Schewe [12]; however, due to page restrictions, we cannot go into detail here.

## 4. Critical Graphs

Strategy iteration performs usually very well on randomly generated games, game families considered as difficult, as well as families based on practical examples. However, there are structures confusing it. We will use two of them to finally construct a binary counter, leading to a family of games being of quadratic size requiring an exponential number of iterations. Those two structures will be referred to as *deceleration lanes* and *stubborn cycles*.

The *deceleration lane* is a family of structures that comprise two nodes, $s$ and $x$, having outgoing edges to the rest of the game, a lane of nodes, $a_k, \ldots, a_0, c$ having incoming edges from the rest of the game, an internal parallel lane of nodes, $b_k, \ldots, b_0, d$ and finally an internal node $t$. See Figure 1 for an example of a deceleration lane.

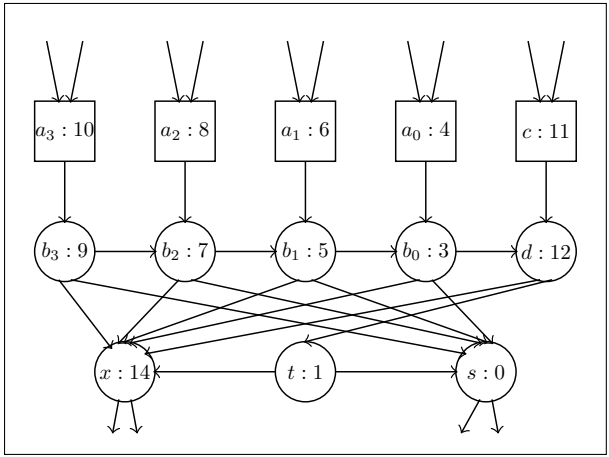

**Figure 1. A Deceleration Lane**

The initial setting for a deceleration lane would be a strategy that maps all player 0 nodes to $x$. Following a run of the strategy improvement algorithm on the whole graph, consider a setting in which the valuation of $x$ remains greater than the one of $s$. In each such iteration, only one edge of the deceleration lane is a proper improvement edge: At first, the edge from $b_0$ to $d$, then the edge from $b_1$ to $b_0$ etc.

At the same time, after updating to the improvement edge, there is always a new node - meaning one in each iteration - accessible from the outside in the deceleration lane that has the highest valuation. In the beginning, the best accessible node is $c$, then $a_0$ and after that $a_1$ etc. To summarize these two properties:

- It takes time linear in its size to be complete: Starting with the initial strategy, each iteration updates one strategy decision.

- It comprises a new best-valuated node in each iteration: Namely the node using the updated strategy edge.

There is another important feature of deceleration lanes: Whenever the valuation of $s$ gets better than the one of $x$, all nodes immediately switch to $s$, except for $d$ (since in this case, the edge from $d$ to $t$ is not an improvement edge).

Therefore, the whole strategy-structure of the deceleration lane can be reset simply be valuating $s$ higher than $x$, even if it is only for the duration of one iteration. After that, the deceleration lane can restructure itself in the way described before.

The awkward construction using the node $t$ to postpone the update of $d$ is not necessary to "fool" the locally improving policy (it suffices to have an edge from $d$ to $s$, omitting $t$ completely); Schewe's globally improving policy, however, requires it. To sum this property up:

- It resets in one step: If there is an event valuating $s$ better than $x$, the whole sub-strategy associated with the deceleration lane immediately gets reset by the improvement policy.

| $\sigma$ | $\Xi_\sigma(s) \prec \Xi_\sigma(x)$ | $t$ | $d$ | $b_0$ | $b_1$ | $b_2$ | $b_3$ | $\mathcal{M}$ |
|---|---|---|---|---|---|---|---|---|
| $\sigma_0$ | $Yes$ | $x$ | $x$ | $x$ | $x$ | $x$ | $x$ | X |
| $\sigma_1$ | $Yes$ | $x$ | $x$ | $d$ | $x$ | $x$ | $x$ | 0 |
| $\sigma_2$ | $Yes$ | $x$ | $x$ | $d$ | $b_0$ | $x$ | $x$ | 1 |
| $\sigma_3$ | $Yes$ | $x$ | $x$ | $d$ | $b_0$ | $b_1$ | $x$ | 2 |
| $\sigma_4$ | $Yes$ | $x$ | $x$ | $d$ | $b_0$ | $b_1$ | $b_2$ | 3 |
| $\sigma_5$ | $No$ | $s$ | $x$ | $s$ | $s$ | $s$ | $s$ | S |
| $\sigma_6$ | $No$ | $s$ | $t$ | $s$ | $s$ | $s$ | $s$ | T |
| $\sigma_7$ | $Yes$ | $x$ | $x$ | $x$ | $x$ | $x$ | $x$ | X |
| $\vdots$ | $\vdots$ | $\vdots$ | $\vdots$ | $\vdots$ | $\vdots$ | $\vdots$ | $\vdots$ | |

**Figure 2. Activity of a Deceleration Lane**

Figure 2 illustrates the update activity of the deceleration lane: The first column shows the sequence of strategies associated with a run of the strategy iteration on a game containing the deceleration lane, the second column shows which of the two nodes having edges leading out of the lane has a better valuation, the last column assigns a macroscopic title to each line (we will refer to these later) and the other columns show the strategy decisions of the current strategy. Note that the external event that resets the lane takes two iterations in this example.

A deceleration lane is used to absorb the update activity of other nodes in such a way that wise strategy updates are postponed. A simple scenario would be a cycle of a player 0 and a player 1 node: Assume that a wise strategy for player 0 is to move to the player 1 node s.t. player 1 is forced to leave the cycle; see Figure 3 for an example of such a situation.

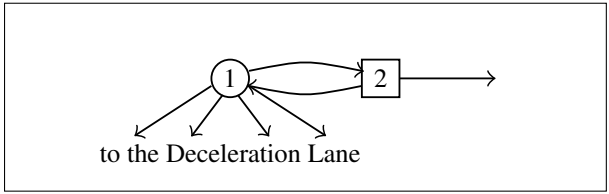

**Figure 3. Usage Example**

Player 0 will be updating to this edge iff there is no edge leading out of the cycle that is better than the edge used before. A deceleration lane thus is a device to fulfill these needs with the addition to be reusable due to its ability to reset itself.

Although such a simple cycle can be used to fool the locally optimizing policy in combination with a deceleration lane, this is not the case with the globally optimizing policy. We are not able to go into detail due to page restrictions; we phrase the problem therefore that way: If we want to postpone player 0 updates closing a cycle, we basically need to make sure that this is at least impossible to be done in one iteration.

This can be achieved by creating a cycle consisting of more than one player 0 node s.t. there is always at least one edge belonging to the cycle that is no improvement edge (as long as closing the cycle should be postponed). A *stubborn cycle* uses three player 0 nodes, say $e$, $f$ and $g$ and one player 1 node, say $h$, to form a cycle.

All three player 0 nodes provide further outgoing edges leading to other structures, for instance to a deceleration lane, in the game graph. We say a stubborn cycle is $\sigma$-*closed* (w.r.t. a strategy $\sigma$) iff all player 0 edges belonging to the cycle are conforming to $\sigma$. See Figure 4 for an example of a stubborn cycle.

To postpone the closing of the cycle, one has to create a setting in which the best outgoing improvement has to change in each iteration, e.g. in the first iteration, the best edge goes from $e$ to somewhere else, in the second iteration, the best edge goes from $f$ to somewhere else, in the third iteration, the best edge goes from $g$ to somewhere else and thereafter the best edge goes again from $e$ to somewhere else etc.

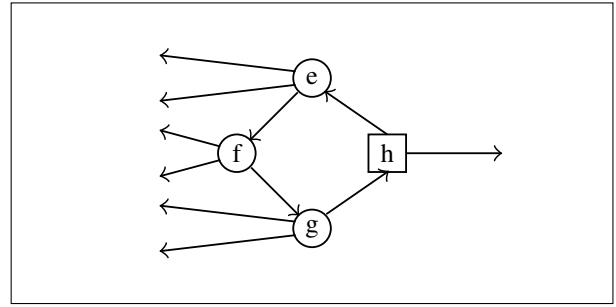

**Figure 4. A Stubborn Cycle**

By providing such a setting, the stubborn cycle maintains the invariant that one of the three player 0 edges building the cycle is a non-improvement edge. In each iteration, the single edge belonging to the cycle is improved to lead out of the cycle while one of the currently two edges leading out of the cycle is improved to lead into the cycle.

A stubborn cycle can therefore be combined with deceleration lane as follows: $e$ is connected to the 0th, 3rd, 6th, etc. entry node of the deceleration lane, $f$ is connected to the 2nd, 5fth etc. entry node and $g$ is connected to the 1st, 4th, etc. entry node.

Figure 5 illustrates the update activity of a stubborn cycle: The first column shows the sequence of strategies associated with a run of the strategy iteration on a game containing the stubborn cycle, the second column shows the associated valuation ranking of three nodes $o_e$, $o_f$ and $o_g$ that can be reached by the respective outgoing edges of the player 0 nodes belonging to the stubborn cycle (in combination with a deceleration lane, there is usually more than one node of the lane that can be reached by a player 0 node of the stubborn cycle). The other columns denote the status of the respective edge w.r.t. the current strategy: Strategy (S), Improvement Edge (I) and Non-Improvement Edge (N).

| $\sigma$ | $\Xi_\sigma$ | $e \to$ | | $f \to$ | | $g \to$ | |
|---|---|---|---|---|---|---|---|
| | | $f$ | $o_e$ | $g$ | $o_f$ | $h$ | $o_g$ |
| $\sigma_0$ | $\Xi(o_g) \prec \Xi(o_f) \prec \Xi(o_e)$ | S | I | N | S | I | S |
| $\sigma_1$ | $\Xi(o_f) \prec \Xi(o_e) \prec \Xi(o_g)$ | N | S | I | S | S | I |
| $\sigma_2$ | $\Xi(o_e) \prec \Xi(o_g) \prec \Xi(o_f)$ | I | S | S | I | N | S |
| $\sigma_3$ | $\Xi(o_g) \prec \Xi(o_f) \prec \Xi(o_e)$ | S | I | N | S | I | S |
| ⋮ | ⋮ | ⋮ | ⋮ | ⋮ | ⋮ | ⋮ | ⋮ |

**Figure 5. Activity of a Stubborn Cycle**

Again, we note that we only use stubborn cycles instead of simple cycles in order to create a worst-case family that works both for the locally improving policy and for the globally improving policy.

## 5. Super-Polynomial Lower Bound

We present a family of parity games of quadratic size requiring exponentially many iterations to be solved by the strategy improvement algorithm. The games are denoted by $\mathcal{G}_n$. The set of nodes are $\mathcal{V}_n := \mathcal{V}_n^0 \cup \mathcal{V}_n^1$, where $\mathcal{V}_n^i$ denote the sets of nodes owned by player $i$:

$$\mathcal{V}_n^0 := \{y, d, s, t, x, w, b_0, \ldots, b_{3n},$$
$$e_0, \ldots, e_{n-1}, f_0, \ldots, f_{n-1}, g_0, \ldots, g_{n-1},$$
$$l_0, \ldots, l_{n-1}, z_0, \ldots, z_{n-1}\}$$
$$\mathcal{V}_n^1 := \{c, p, q, a_0, \ldots, a_{3n}, h_0, \ldots, h_{n-1},$$
$$k_0, \ldots, k_{n-1}, m_0, \ldots, m_{n-1}\}$$

Please refer to Figure 6 for the priority function and the edges of $\mathcal{G}_n$. The game $\mathcal{G}_2$ is depicted in Figure 7.

| | Node | Priority | Successors |
|---|---|---|---|
| Deceleration Lane | $s$ | $2$ | $\{p\} \cup \{k_j \mid j < n\}$ |
| | $t$ | $3$ | $\{x, s\}$ |
| | $y$ | $5$ | $\{l_0, w\}$ |
| | $w$ | $7$ | $\{p, y\} \cup \{l_j \mid 0 < j < n\}$ |
| | $b_0$ | $6n + 9$ | $\{s, x, d\}$ |
| | $b_{i>0}$ | $6n + 2i + 9$ | $\{s, x, b_{i-1}\}$ |
| | $a_i$ | $6n + 2i + 10$ | $\{b_i\}$ |
| | $c$ | $12n + 11$ | $\{d\}$ |
| | $d$ | $12n + 12$ | $\{t, x\}$ |
| | $x$ | $12n + 14$ | $\{w, y\}$ |
| Stubborn C. | $e_i$ | $6i + 9$ | $\{s, f_i, c\} \cup$ $\{a_{3j+2} \mid j \leq i\}$ |
| | $f_i$ | $6i + 11$ | $\{g_i\} \cup \{a_{3j+1} \mid j \leq i\} \cup$ $\{k_j \mid j < n\}$ |
| | $g_i$ | $6i + 13$ | $\{h_i\} \cup \{a_{3j} \mid j \leq i + 1\}$ |
| | $h_i$ | $6i + 14$ | $\{e_i, m_i\}$ |
| C.-assoc. | $l_i$ | $6i + 10$ | $\{k_i, z_i\}$ |
| | $z_i$ | $12n + 4i + 15$ | $\{p\} \cup \{l_j \mid i < j < n\}$ |
| | $k_i$ | $12n + 4i + 17$ | $\{h_i\}$ |
| | $m_i$ | $12n + 4i + 18$ | $\{z_i\}$ |
| End C. | $q$ | $1$ | $\{q\}$ |
| | $p$ | $16n + 16$ | $\{q\}$ |

**Figure 6. The Game $\mathcal{G}_n$**

**Corollary 6.** *The game $\mathcal{G}_n$ has $14 \cdot n + 11$ nodes, $3 \cdot n^2 + 28 \cdot n + 17$ edges and $16 \cdot n + 16$ as highest priority. In particular, $|\mathcal{G}_n| = \mathcal{O}(n^2)$.*

First, we note that every approach trying to construct a game family of polynomial size requiring exponentially many iterations to be solved, needs to focus on the second component of game valuations: There are only linearly many different values for the first and third component while there are exponentially many for the second.

The games $\mathcal{G}_n$ therefore are designed in such a way that the first and the third component of all occurring game valuations in a run of the algorithm on $\mathcal{G}_n$ remain unaltered: Following the initial strategy (selecting the successor with the greatest reward), it is not hard to see that the worst (w.r.t. player 0) dominating cycle node that can be reached is $q$.

Additionally note that $\mathcal{G}_n$ is completely won by player 1, therefore there is no dominating cycle node that could be reached during the iteration process that is better than $q$ (since $q$ is the node with the greatest negative reward in $\mathcal{G}_n$). Due to the fact that all other nodes have greater priorities than $q$, the third component of the valuations (basically measuring the number of less relevant nodes) remains useless. Thus, we will only focus on the paths leading to $q$.

The basic idea is to create a binary counter in $\mathcal{G}_n$: To formalise the state of an $n$-bit counter, we use $n$-tuples $\alpha = (\alpha_{n-1} \ldots \alpha_0) \in \{0, 1\}^n$ where $\alpha_0$ denotes the lowest bit in $\alpha$ and $\alpha_{n-1}$ denotes the highest bit. The lowest and highest values in $\{0, 1\}^n$ will be abbreviated by $\mathbf{0}_n := (0 \ldots 0)$ and $\mathbf{1}_n := (1 \ldots 1)$.

Generally, the games $\mathcal{G}_n$ consist of a long deceleration lane (being built from the nodes $a_i$, $b_i$, $c$, $d$, $t$, $x$, and $s$), $n$ stubborn cycles (being built from the nodes $e_i$, $f_i$, $g_i$ and $h_i$) connected to the deceleration lane, cycle-associated structures ($k_i$, $l_i$, $m_i$ and $z_i$), the cycle in which all valuation-occurring paths end ($p$ and $q$) as well as two additional nodes ($y$ and $w$) being associated with the deceleration lane.

The sequence of strategies that is associated with a run on $\mathcal{G}_n$ can be separated into three phases: An initialization phase (of length $\mathcal{O}(1)$), a counting phase (of length $2^{\mathcal{O}(n)}$) and a finalization phase (of length $\mathcal{O}(n)$). The following clarifications will be focusing on the counting phase. The counter starts in the counting phase with the lowest bit set and finishes with all bits set except the lowest bit.

The binary counter is implemented by the sequence of stubborn cycles. Let $\sigma$ be a fixed strategy occurring during the counting phase. A $\sigma$-closed stubborn cycle (forcing player 1 to move out of the cycle) should be considered as a set bit in the counter while a $\sigma$-open stubborn cycle should be considered as bit which is not set.

Intuitively, the strategy $\sigma$ can be described as follows: All nodes belonging to open stubborn cycles follow edges to the deceleration lane; the deceleration lane follows edges to the entry point of the lowest closed stubborn cycle which itself follows edges to the next closed stubborn cycle (via the cycle-associated structures) etc.; the last closed stubborn cycle follows edges directly leading to the end of the game.

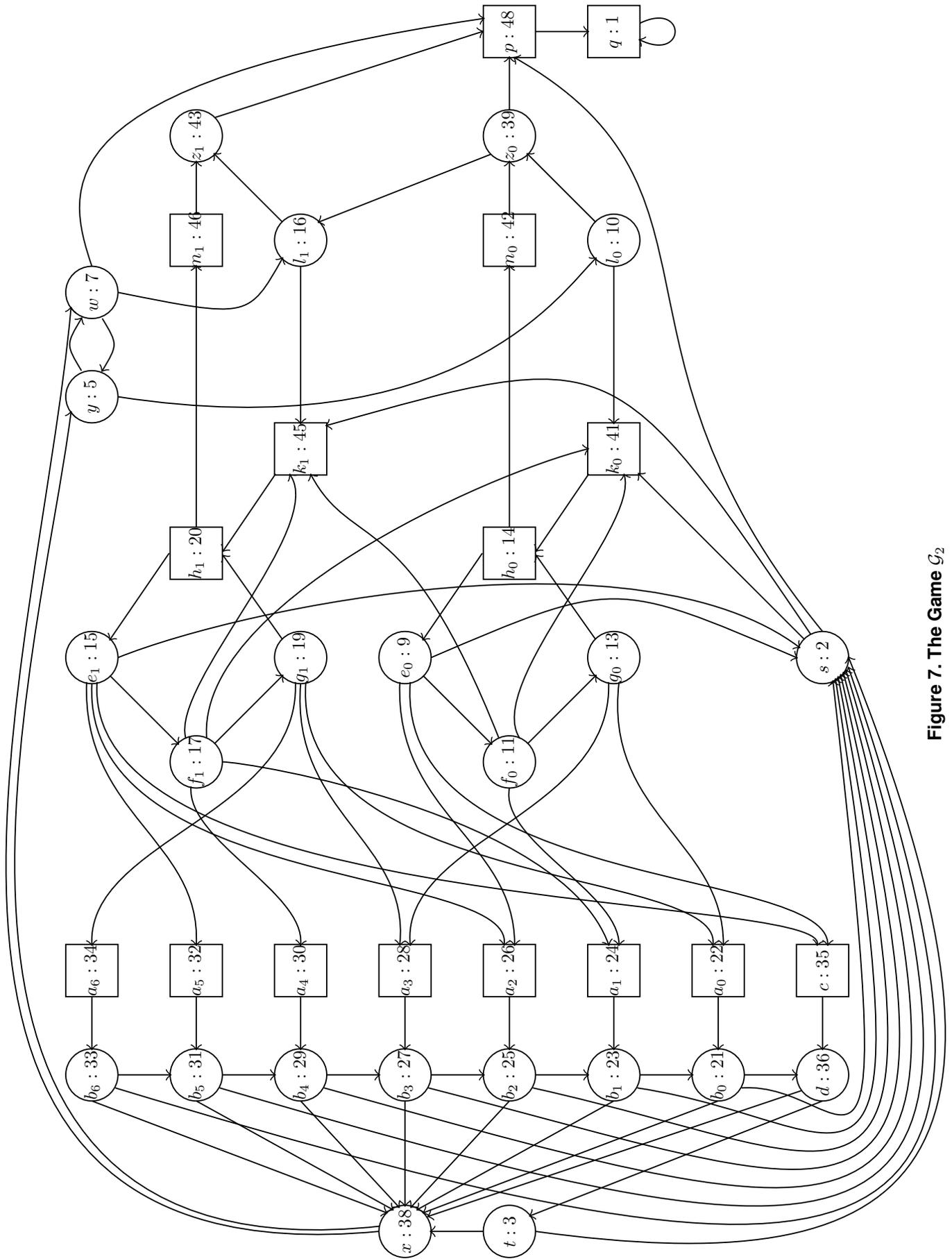

Figure 7. The Game $\mathcal{G}_2$

Let $\alpha \in \{0,1\}^n$ be the bit-state associated with $\sigma$ and assume that $\alpha \neq \mathbf{0}_n, \mathbf{1}_n$.

Since we want to implement a counter in the game, the next bit-state that $\alpha$ changes to during the subsequent iterations should be $\alpha^+$ (denoting the bit-increment of $\alpha$; similarly, $\alpha^-$ denotes the bit-decrement of $\alpha$).

This, basically, can be achieved as follows: All player 0 nodes belonging to the stubborn cycles have edges pointing out of the cycles to the deceleration lane, whereas stubborn cycles associated with bit $i$ only have edges up to $a_{3i+3}$.

All stubborn cycles whose bits are set are connected using the $z_i$ nodes. The most profitable path through the cycle area therefore starts with the lowest cycle whose bit is set, running through all other closed stubborn cycles, finally ending in $p$.

The number of iterations that the counter needs in order to increment depends on the lowest 0-bit, denoted by $\mu_\alpha := \min\{j \mid \alpha_j = 0\}$ (keep in my mind that we assume $\alpha \neq \mathbf{1}_n$), and is around $\gamma_\alpha := 3 \cdot \mu_\alpha + 7$. We will call a sequence of subsequently improved strategies that are associated with one $\alpha$ a *round*.

The iterations of a round are mainly absorbed by the update activity of the deceleration lane: At the beginning of a round, all player 0 nodes of the deceleration lane, $b_0, \ldots, b_{3n}$, point to the starting node $s$, after that to $x$ and then subsequently to the respective lower neighbour, i.e. $b_0$ to $d$, $b_1$ to $b_0$ etc.; we call a node $a_i$ $\sigma$-*chaining node* iff its direct successor $b_i$ already moves to its lower neighbour, i.e. $\sigma(b_i) = b_{i-1}$ if $i > 0$ and $\sigma(b_i) = d$ if $i = 0$.

During a round, all closed stubborn cycles remain closed until the end of the round since it is more profitable to keep a cycle closed than to also move to the deceleration lane. All open stubborn cycles basically point to the highest $a_i$ chaining node by the current strategy.

Following the update activity of the deceleration lane, all open stubborn cycles are absorbed by updating themselves to always reach the highest chaining node.

The first open stubborn cycle, i.e. the one associated with $i := \mu_\alpha$, only follows the update activity of the deceleration lane up to $a_{3i+3}$. After reaching that point, the next iterations lead to a closing of the $i$-th stubborn cycle. Additionally, the strategy decision for $k_i$ is improved to move the closed stubborn cycle itself. Since all lower $f_j$ nodes have an edge leading to $k_i$, the next update that happens is the opening of all lower stubborn cycles, simulating the correct increment of $\alpha$.

The cycle access node $s$ also updates immediately after closing the stubborn cycle to move there, leading to a reset of the whole deceleration lane. The node $x$ also updates to move there (over the node $y$ or $w$ resp.) one iteration after that s.t. the next round is about to begin.

From a macroscopic point of view, a strategy $\sigma$ occurring during the counting phase can be roughly described by the macroscopic state $\mathcal{M}$ of the deceleration lane and the bit-states of the stubborn cycles. Figure 8 illustrates the counting phase of $\mathcal{G}_3$.

| $\sigma$ | $\alpha_2$ | $\alpha_1$ | $\alpha_0$ | $\mathcal{M}$ |
|---|---|---|---|---|
| $\sigma_0$ | 0 | 0 | 1 | T |
| $\sigma_1$ | 0 | 0 | 1 | X |
| $\sigma_2$ | 0 | 0 | 1 | 0 |
| $\sigma_3$ | 0 | 0 | 1 | 1 |
| ⋮ | ⋮ | ⋮ | ⋮ | ⋮ |
| $\sigma_{11}$ | 0 | 0 | 1/0 | 9 |
| $\sigma_{12}$ | 0 | 1 | 0 | S |
| $\sigma_{13}$ | 0 | 1 | 0 | T |
| $\sigma_{14}$ | 0 | 1 | 0 | X |
| $\sigma_{15}$ | 0 | 1 | 0 | 0 |
| $\sigma_{16}$ | 0 | 1 | 0 | 1 |
| ⋮ | ⋮ | ⋮ | ⋮ | ⋮ |
| $\sigma_{21}$ | 0 | 1 | 0 | 6 |
| $\sigma_{22}$ | 0 | 1 | 1 | S |
| ⋮ | ⋮ | ⋮ | ⋮ | ⋮ |
| $\sigma_{70}$ | 1 | 1 | 0 | 6 |
| $\sigma_{71}$ | 1 | 1 | 1 | S |

**Figure 8. Counting Phase**

The sequence of strategies associated with a run of the improvement algorithm (using the locally optimizing policy) on $\mathcal{G}_n$ can be decomposed into three phases:

1. An initialization phase that always requires 11 strategies: $\sigma^0_{(n,\beta)}$ where $-2 \leq \beta \leq 8$

2. A counting phase in which flipping a bit requires $\gamma_\alpha + 3$ strategies for a given counter state $\alpha$: $\sigma^1_{(n,\alpha,\beta)}$ where $\alpha \neq \mathbf{0}_n, \mathbf{1}_n$ and $-2 \leq \beta \leq \gamma_\alpha$

3. A finalization phase that requires $3 \cdot n + 3$ strategies: $\sigma^2_{(n,\beta)}$ where $-2 \leq \beta \leq 3 \cdot n$

The definition of $\sigma^0_{(n,\beta)}, \sigma^1_{(n,\alpha,\beta)}$ and $\sigma^2_{(n,\beta)}$ has been put into the appendix. The following lemma shows in which order these strategies are being chosen in a run of the algorithm on $\mathcal{G}_n$.

**Lemma 7.** *Let $n > 0$. We apply the following notation for every 0-strategy $\sigma$ compliant with $\mathcal{G}_n$: $\sigma' := \mathcal{I}^{\mathrm{loc}}_{\mathcal{G}_n}(\sigma)$. Then the following holds:*

1. $(\iota_{\mathcal{G}_n})' = \sigma^0_{(n,-2)}$

2. $(\sigma^0_{(n,\beta)})' = \sigma^0_{(n,\beta+1)}$ *for every* $-2 \leq \beta < 8$

3. $(\sigma^0_{(n,8)})' = \sigma^1_{(n,\mathbf{0}^+_n,-2)}$

4. $(\sigma^1_{(n,\alpha,\beta)})' = \sigma^1_{(n,\alpha,\beta+1)}$ *for every* $\alpha \neq \mathbf{0}_n, \mathbf{1}_n$ *and* $-2 \leq \beta < \gamma_\alpha$

5. $(\sigma^1_{(n,\alpha,\gamma_\alpha)})' = \sigma^1_{(n,\alpha^+,-2)}$ *for every* $\alpha \neq \mathbf{0}_n, \mathbf{1}_n^-, \mathbf{1}_n$

6. $(\sigma^1_{(n,\mathbf{1}_n^-,\gamma_{\mathbf{1}_n^-})})' = \sigma^2_{(n,-2)}$

7. $(\sigma^2_{(n,\beta)})' = \sigma^2_{(n,\beta+1)}$ *for every* $-2 \leq \beta < 3 \cdot n$

8. $(\sigma^2_{(n,3\cdot n)})' = \sigma^2_{(n,3\cdot n)}$

We omit the easy but tedious proofs due to page restrictions here. Technically, one simply needs to compute the result of the improvement policy in each step. We show one tiny part of the proof in the appendix.

By induction on $n$ using the former lemma we can conclude:

**Theorem 8.** *Let* $n > 0$. *The strategy improvement algorithm requires* $13 \cdot 2^n - 9$ *iterations on* $\mathcal{G}_n$ *using the locally optimizing improvement policy.*

Hence, the strategy iteration using the locally optimizing policy requires at least super-polynomial time in the worst case (since each iteration requires polynomial time and $\mathcal{G}_n$ is of quadratic size; see Lemma 2, 5 and Corollary 6).

We implemented an open-source parity game solver platform, the PGSOLVER Collection [3], that particularly contains implementations of the strategy iteration due to Vöge and Jurdziński [16] as well as the variant by Schewe [12]. Benchmarking both algorithms with $\mathcal{G}_n$ results in exponential run-time behaviour as can be seen in Figure 9 (note that the time-axis has logarithmic scale).

## 6. Conclusion

Vöge mentions in his PhD thesis [15] that it is probably much more convenient for strategy improvement algorithms to be performed on games with an outgoing edge degree limited by two. A simple transformation of the family of games presented here resulting in a family of games with out-degree limited by two also requires super-polynomial time.

There are other possibilities to select the initial strategy. Randomizing the initial strategy, for instance, is another popular choice: we note without proof that starting with a randomized strategy, the expected number of iterations on $\mathcal{G}_n$ is also super-polynomial.

By applying some simple transformations on $\mathcal{G}_n$ (such as alternating transformation etc.), these games can also be used to show that both variants ([12] and [10]) of Schewe's globally optimizing technique require the same number of iterations to be solved, and therefore also super-polynomial time.

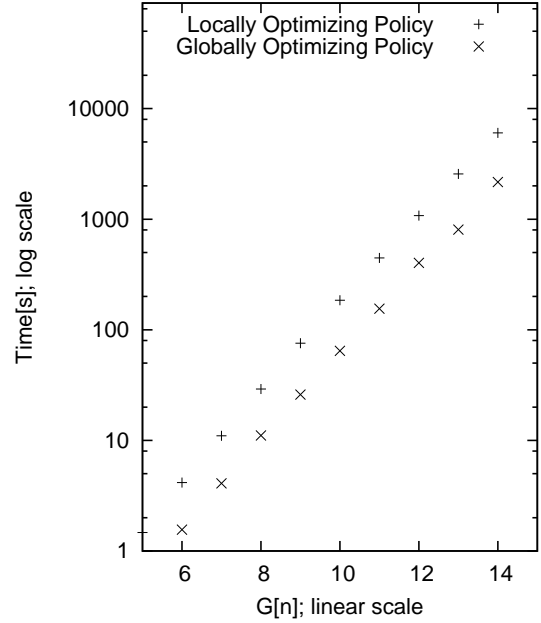

**Figure 9. The Benchmark**

Although there are many preprocessing techniques that could be used to simplify the family of games presented here - e.g. decomposition into strongly connected components, compression of priorities, direct-solving of simple cycles, etc. - such procedures cannot be implemented to fix the bad performance of the strategy iteration on these games since all known preprocessing techniques can be fooled quite easily without really touching the inner structure of the game.

The same applies to simultaneous solving using different algorithms due to the fact that it is not very complicated to combine different worst-case games in such a way that each algorithm that tries to solve the whole game is slowed down by the part that belongs to its worst-case example.

Parity games are widely believed to be solvable in polynomial time, yet there is no algorithm known that is performing better than super-polynomially. Vöge presented his strategy iteration technique in his PhD thesis eight years ago, and this class of solving procedures is generally supposed to be the best candidate to give rise to an algorithm that solves parity games in polynomial time since then. Unfortunately the two most obvious improvement policies, namely the locally and the globally optimizing technique, are not capable of doing so.

We think that the strategy iteration still is a promising candidate for a polynomial time algorithm, although it is possibly necessary to alter more of it than just the improvement policy. The main problem of the algorithm (and the policies) is that reoccurring substructures are not handled in such a way that a combination of edges that was profitable before is applied again. The reason is that possibly not all edges belonging to that profitable combination are improvement edges, hence that combination cannot be selected in a single improvement step.

Therefore we believe that it would be an interesting approach to add some kind of memorization of profitable substructures that can be applied as a whole under certain conditions that are weaker than requiring all edges of the substructure to be improvement edges but strong enough to ensure the soundness of the algorithm.

**Acknowledgements.** I am indebted to Martin Lange and Martin Hofmann for their guidance and numerous inspiring discussions on the subject.

## A. Appendix

We will apply the following additional notations for $\alpha \in \{0,1\}^n$:

- $\nu_\alpha := \max\{j \leq n \mid \forall k < j.\alpha_k = 0\}$
- $\alpha|_j := (\alpha_{n-1} \ldots \alpha_{j+1} 0 \ldots 0)$

The initialization strategy family $\sigma^0_{(n,\beta)}$ where $-2 \leq \beta \leq 8$ is defined as follows: $\sigma^0_{(n,\beta)} :=$

$$\begin{cases} b_0 & \mapsto \begin{cases} s & \text{if } \beta = -1, 8 \\ x & \text{if } \beta = -2 \\ d & \text{otherwise} \end{cases} \\ b_{j>0} & \mapsto \begin{cases} s & \text{if } \beta = -1, 8 \\ x & \text{if } \beta \neq -1 \wedge \beta < j \\ b_{j-1} & \text{otherwise} \end{cases} \\ l_j & \mapsto \begin{cases} k_j & \text{if } j = 0 \wedge \beta \geq 7 \\ z_j & \text{otherwise} \end{cases} \\ s & \mapsto \begin{cases} k_0 & \text{if } \beta \geq 7 \\ p & \text{otherwise} \end{cases} \\ t & \mapsto \begin{cases} s & \text{if } \beta = -1, 8 \\ x & \text{otherwise} \end{cases} \\ x & \mapsto \begin{cases} y & \text{if } \beta = -2 \\ w & \text{otherwise} \end{cases} \\ y & \mapsto \begin{cases} l_0 & \text{if } \beta = -2, 8 \\ w & \text{otherwise} \end{cases} \\ e_j & \mapsto \begin{cases} a_{3j+2} & \text{if } \beta = -2 \\ s & \text{if } \beta = -1 \vee \\ & \quad (\beta = 8 \wedge j > 0) \\ c & \text{if } \beta = 0, 1 \\ a_2 & \text{if } \beta = 3, 4 \\ a_5 & \text{if } \beta = 6, 7 \wedge j > 0 \\ f_j & \text{otherwise} \end{cases} \\ f_j & \mapsto \begin{cases} k_0 & \text{if } \beta = 7, 8 \wedge j > 0 \\ a_{3j+1} & \text{if } \beta = -2 \\ a_1 & \text{if } \beta = 2, 3 \\ a_4 & \text{if } \beta = 5, 6 \wedge j > 0 \\ g_j & \text{otherwise} \end{cases} \\ g_j & \mapsto \begin{cases} a_{3j+3} & \text{if } \beta < 1 \\ a_0 & \text{if } \beta = 1, 2 \\ a_3 & \text{if } \beta = 4, 5 \\ a_6 & \text{if } \beta \geq 7 \wedge j > 0 \\ h_j & \text{otherwise} \end{cases} \\ d & \mapsto x \\ w & \mapsto p \\ z_j & \mapsto p \end{cases}$$

The counting strategy family $\sigma^1_{(n,\alpha,\beta)}$ where $\alpha \neq \mathbf{0}_n, \mathbf{1}_n$ and $-2 \leq \beta \leq \gamma_\alpha$ is defined as follows: $\sigma^1_{(n,\alpha,\beta)} :=$

$$\begin{cases} b_0 & \mapsto \begin{cases} s & \text{if } \beta = -2, \gamma_\alpha \\ x & \text{if } \beta = -1 \\ d & \text{otherwise} \end{cases} \\ b_{j>0} & \mapsto \begin{cases} s & \text{if } \beta = -2, \gamma_\alpha \\ x & \text{if } -2 < \beta < j, \gamma_\alpha \\ b_{j-1} & \text{otherwise} \end{cases} \\ d & \mapsto \begin{cases} t & \text{if } \beta = -2 \\ x & \text{otherwise} \end{cases} \\ l_j & \mapsto \begin{cases} k_j & \text{if } \alpha_j = 1 \vee \\ & \quad (\beta \geq \gamma_\alpha - 1 \wedge j = \mu_\alpha) \\ z_j & \text{otherwise} \end{cases} \\ s & \mapsto \begin{cases} k_{\nu_\alpha} & \text{if } \beta < \gamma_\alpha - 1 \\ k_{\mu_\alpha} & \text{otherwise} \end{cases} \\ t & \mapsto \begin{cases} s & \text{if } \beta = -2, \gamma_\alpha \\ x & \text{otherwise} \end{cases} \\ w & \mapsto \begin{cases} y & \text{if } \alpha_0 = 1 \wedge \beta < \gamma_\alpha \\ l_{\mu_\alpha} & \text{if } \alpha_0 = 1 \wedge \beta = \gamma_\alpha \\ l_{\nu_\alpha} & \text{otherwise} \end{cases} \\ x & \mapsto \begin{cases} y & \text{if } \alpha_0 = 1 \\ w & \text{otherwise} \end{cases} \\ y & \mapsto \begin{cases} l_0 & \text{if } \alpha_0 = 1 \vee \beta = \gamma_\alpha \\ w & \text{otherwise} \end{cases} \\ z_j & \mapsto \begin{cases} l_{\mu_\alpha} & \text{if } \beta = \gamma_\alpha \wedge j < \mu_\alpha \\ l_{\nu_{\alpha|_j}} & \text{if } j < \nu_\alpha|_j < n \wedge \\ & \quad (\beta < \gamma_\alpha \vee \mu_\alpha \leq j) \\ p & \text{otherwise} \end{cases} \\ e_j & \mapsto \begin{cases} c & \text{if } \alpha_j = 0 \wedge \beta = -1, 0 \\ s & \text{if } \alpha_j^+ = 0 \wedge \beta = \gamma_\alpha \\ a_{3j+2} & \text{if } \alpha_j = 0 \wedge \beta = -2 \\ a_{\beta-1} & \text{if } \beta > 0 \wedge \beta \equiv_3 0 \wedge \alpha_j = 0 \wedge \\ & \quad (\beta < \gamma_\alpha - 1 \vee \mu_\alpha < j) \\ f_j & \text{otherwise} \end{cases} \\ f_j & \mapsto \begin{cases} a_{3j+1} & \text{if } \beta = -2 \wedge \alpha_j = 0 \wedge (\alpha_j^- = 1 \vee \\ & \quad (\alpha_j^- = 0 \wedge \gamma_{\alpha^-} \leq 3j+1)) \\ a_{\beta-1} & \text{if } \beta > 0 \wedge \beta \equiv_3 2 \wedge \\ & \quad \alpha_j = 0 \wedge (\beta < \gamma_\alpha - 2 \vee \mu_\alpha < j) \\ k_{\mu_\alpha} & \text{if } \alpha_j^+ = 0 \wedge \beta \geq \gamma_\alpha - 1 \\ g_j & \text{otherwise} \end{cases} \\ g_j & \mapsto \begin{cases} a_{3j+3} & \text{if } \alpha_j = 0 \wedge \beta < 0 \\ a_{\beta-1} & \text{if } \beta > 0 \wedge \beta \equiv_3 1 \wedge \alpha_j = 0 \wedge \\ & \quad (\beta < \gamma_\alpha \vee \mu_\alpha < j) \\ h_j & \text{otherwise} \end{cases} \end{cases}$$

The finalization strategy family $\sigma^2_{(n,\beta)}$ where $-2 \leq \beta \leq 3 \cdot n$ is defined as follows: $\sigma^2_{(n,\beta)} :=$

$$\begin{cases} b_0 & \mapsto \begin{cases} s & \text{if } \beta = -2 \\ x & \text{if } \beta = -1 \\ d & \text{otherwise} \end{cases} \\ b_{j>0} & \mapsto \begin{cases} s & \text{if } \beta = -2 \\ x & \text{if } -2 < \beta < j \\ b_{j-1} & \text{otherwise} \end{cases} \\ z_j & \mapsto \begin{cases} p & \text{if } j = n-1 \\ l_{j+1} & \text{otherwise} \end{cases} \\ d & \mapsto \begin{cases} t & \text{if } \beta = -2 \\ x & \text{otherwise} \end{cases} \\ t & \mapsto \begin{cases} s & \text{if } \beta = -2 \\ x & \text{otherwise} \end{cases} \\ l_j & \mapsto k_j \\ s & \mapsto k_0 \\ w & \mapsto y \\ x & \mapsto y \\ y & \mapsto l_0 \\ e_j & \mapsto f_j \\ f_j & \mapsto g_j \\ g_j & \mapsto h_j \end{cases}$$

As an example on how to prove the obligations of Lemma 7, we show the following:

**Example 9.** *Let $n > 0$, $\alpha \neq \mathbf{0}_n, \mathbf{1}_n^-, \mathbf{1}_n$ and $0 < \beta < \gamma_\alpha - 2$. Then:*

$$\mathcal{I}^{\text{loc}}_{\mathcal{G}_n}(\sigma^1_{(n,\alpha,\beta)}) = \sigma^1_{(n,\alpha,\beta+1)}$$

*Proof.* Let $n > 0$, $\alpha \neq \mathbf{0}_n, \mathbf{1}_n^-, \mathbf{1}_n$, $0 < \beta < \gamma_\alpha - 2$, $\sigma := \sigma^1_{(n,\alpha,\beta)}$, $\sigma^* := \sigma^1_{(n,\alpha,\beta+1)}$, $\sigma' := \mathcal{I}^{\text{loc}}_{\mathcal{G}_n}(\sigma)$ and $\Xi := \Xi_\sigma$.

Due to the fact that the first component of $\Xi(v)$ is $q$ and the third component is irrelevant for all $v$ (since there are no nodes being less relevant than $q$), we omit the first and third component here completely and identify $\Xi(v)$ with the path component of $\Xi(v)$.

First, we observe that

$$\begin{aligned} \Xi(l_j) \prec \Xi(l_i) &\iff \\ \alpha|_{j-1} < \alpha|_{i-1} &\vee (\alpha|_{j-1} = \alpha|_{i-1} \wedge \\ (\alpha_j < \alpha_i &\vee (\alpha_j = \alpha_i \wedge i < j))) \end{aligned} \quad (1)$$

holds. By 1, we can therefore conclude that $\sigma'(z_i) = \sigma^*(z_i)$ for all $i$. Moreover it is obvious that $\sigma'(y) = \sigma^*(y)$, $\sigma'(w) = \sigma^*(w)$ and thus $\sigma'(x) = \sigma^*(x)$.

Also note that

$$\Xi(\sigma(z_j)) \prec \Xi(l_j) \iff \alpha_j = 1 \quad (2)$$

holds. Therefore, we can conclude that $\sigma'(l_i) = \sigma^*(l_i)$ for all $i$.

Consider that

$$\forall j \neq \nu_\alpha : \Xi(k_j) \prec \Xi(k_{\nu_\alpha}) \quad (3)$$

and therefore $\sigma'(s) = \sigma^*(s)$.

Now note that $s$ (directly) and $x$ (over $y$ or $w$ and $l_{\nu_\alpha}$) both lead into $k_{\nu_\alpha}$ by $\sigma$. Hence $\Xi(s) \prec \Xi(x)$, thus $\sigma'(t) = \sigma^*(t)$ and $\sigma'(d) = \sigma^*(d)$. We can conclude that also $\sigma'(b_j) = \sigma^*(b_j)$ holds for all $j$.

Finally we need to investigate the update activity of the stubborn cycles. It easy to see that for all closed stubborn cycles, i.e. $\sigma(e_i) = f_i$, $\sigma(f_i) = g_i$ and $\sigma(g_i) = h_i$, holds that they remain closed.

For all open stubborn cycles $i$ the following holds. Assume that $\beta \equiv_3 0$ (the other two cases $\beta \equiv_3 1$ and $\beta \equiv_3 2$ are similar). By definition of $\sigma$ it holds that $\sigma(e_i) = a_{\beta-1}$, $\sigma(f_i) = g_i$ and $\sigma(g_i) = h_i$.

Now note that $\Xi(a_j) \prec \Xi(a_\beta)$ for all $j \neq \beta$. Therefore $\sigma'(e_i) = f_i = \sigma^*(e_i)$, $\sigma'(f_i) = g_i = \sigma^*(f_i)$ and $\sigma'(g_i) = a_\beta = \sigma^*(g_i)$. □